\documentclass[prl,aps,showpacs,twocolumn]{revtex4}

\usepackage{graphicx}
\usepackage{braket}
\usepackage{bm}
\usepackage[latin1]{inputenc}

\begin{document}

\title{Quantum Process Tomography of a M$\o$lmer-S$\o$rensen Interaction}

\author{Nir Navon\footnote{These authors contributed equally to this work.}\footnote{Present email and address: nn270@cam.ac.uk, Cavendish Laboratory, University of Cambridge, J.J. Thomson Ave., Cambridge CB3 0HE, United Kingdom}, Nitzan Akerman$^*$, Shlomi Kotler, Yinnon Glickman, and Roee Ozeri}\affiliation{Department of Physics of Complex Systems, Weizmann Institute of Science, Rehovot 76100, Israel}

\date{\today}

\begin{abstract}
We report the quantum process tomography of a M$\o$lmer-S$\o$rensen entangling gate. The tomographic protocol relies on a single discriminatory transition, exploiting excess micromotion in the trap to realize all operations required to prepare all input states and analyze all output states. Using a master-slave diode lasers setup, we demonstrate a two-qubit entangling gate, with a fidelity of Bell state production of 0.985(10). We characterize its $\chi$-process matrix, the simplest for an entanglement gate on a separable-states basis, and we observe that the dominant source of error is accurately modelled by a quantum depolarization channel.
\end{abstract}

\maketitle
The ability to realize and characterize high-fidelity two-qubit gates is central for quantum information science as, together with single-qubit rotations, they constitute the building blocks for quantum computation \cite{barenco1995elementary}. The detailed characterization of these gates is therefore crucial. Quantum Process Tomography (QPT) is an important method to fully characterize linear quantum processes. In particular, QPT of two-qubit entangling gates has been used to characterize CNOT gates in linear-optic \cite{kiesel2005linear}, NMR \cite{childs2001realization}, as well as trapped ions \cite{riebe2006process,home2009complete}, or a square root i-SWAP gate with superconducting qubits \cite{bialczak2010quantum}. In trapped-ions experiments, M$\o$lmer-S$\o$rensen (MS) entangling gates \cite{sorensen1999quantum} have become increasingly popular, both for quantum computation purposes \cite{sackett2000experimental,leibfried2003experimental} and for inducing effective spin-spin couplings that allow to simulate complex quantum many-body hamiltonians from condensed matter physics \cite{kim2009entanglement}. One of its main advantages as compared with other gate protocols is its first-order insensitivity to the phonon occupation number (i.e. temperature of the ion-crystal), which allowed, \emph{inter alia}, the highest entangled state production fidelity reached to date ($0.993(1)$ \cite{benhelm2008towards}), entanglement between ions in thermal motion \cite{kirchmair2009deterministic}, as well as the creation of a maximally entangled state of a large ($N=14$) number of qubits \cite{monz201114}. In this letter, we first implement a new and simple protocol for QPT with trapped ions, which only requires a single discriminatory transition. The scheme is based on inhomogeneous micromotion in the trap that enables addressing single qubits in the chain \cite{turchette1998deterministic,warring2012individual,navon2012single}. Subsequently, we realize the tomographic reconstruction of a M$\o$lmer-S$\o$rensen interaction which, despite its growing importance, has not been process-analyzed yet. 

A quantum process is defined as a completely positive map $\mathcal{E}$ in the space of density matrices. Given a complete set of operators $\{A_i\}$ (such that $\sum_j A^\dag_j A_j=I$), the output state for an arbitrary input state $\rho$ can be written as (for details see for instance \cite{poyatos1997complete,nielsen2010quantum,childs2001realization})
\begin{equation}\label{chiMatrixDef}
\mathcal{E}[\rho]=\sum_{a,b}\chi_{ab}A_a\rho A^\dag_b.
\end{equation}
Here $\{\chi_{ab}\}$ is the \emph{process matrix} (with $4^n\times4^n$ elements for $n$ qubits), which contains the full information on the process $\mathcal{E}$ and is measured by QPT.
A convenient set of input states for the tomography is the product states $\ket{\psi_i}=\ket{\phi_1} \otimes \ket{\phi_2}$, where $\ket{\phi_1},\ket{\phi_2}\in\{\ket{x},\ket{y},\ket{z},\ket{\bar{z}}\}$, which are the one-qubit eigenstates of the Pauli matrices $\{\sigma_x$,$\sigma_y$,$\sigma_z$,$\sigma_z\}$ with eigenvalues $\{1,1,1,-1\}$. Note that, with this choice, entangled states are not used as input states. The measurement basis is conveniently chosen to be $\sigma_i\otimes\sigma_j$ where $i=\{0,x,y,z\}$, and $\sigma_0=I$. However, in the experiment, the detection scheme relies on the statistics of fluorescence photons, which corresponds to the measurement of the expectation value $\braket{\sigma_z\otimes\sigma_z}=\textrm{Tr}[\rho(\sigma_z\otimes\sigma_z)]$. In order to measure the expectation value of $\braket{\sigma_i\otimes\sigma_j}$, we perform additional rotations on the two qubits. In general these rotations require single-qubit addressing capability. For our purpose, a single discriminatory transition
is sufficient for all the required operations.

In our setup, we use $^{88}$Sr$^+$ ions confined in a linear Paul trap \cite{akerman2011quantum}. We work with optical qubits that are encoded in the $\ket{S}=5S_{1/2,+1/2}$ ground state level and in the $\ket{D}=4D_{5/2,+3/2}$ meta-stable level which has a $1/e$ lifetime of $390$ ms \cite{letchumanan2005lifetime}. Coherent manipulation of the qubit state is performed with a narrow linewidth laser at 674 nm which drives an electric-quadrupole transition \cite{DiodePaperInPrep}. The other Zeeman level of the ground state $\ket{S'}=5S_{1/2,-1/2}$, separated by $12.3$ MHz from the $\ket{S}$ level due to a constant magnetic field, is used as auxiliary level in the state detection scheme. Measuring the qubit state is accomplished by counting fluorescence photons on the $5S_{1/2} \rightarrow 5P_{1/2}$ dipole transition with a photomultiplier tube (PMT). We inferred the number of ions in the $\ket{S}$ (bright) state, for each realization, by the number of detected photons. The probabilities, $P_0$, $P_1$, $P_2$, of finding zero, one and two ions in the $\ket{S}$ (bright) state were estimated by the fraction of realizations with the corresponding number of ions inferred in that state.
The discriminatory transition is provided through a micromotion sideband. In an ideal linear Paul trap, the symmetry axis of the trap is also the axis where the RF vanishes, and no excess micromotion is present. However, due to the finite size of the trap, the boundary conditions set by the endcaps leads to an rf leak along this axis, and the region of rf-null is reduced to a point at the center. 
If the two-ion chain is axially aligned so that one ion sits on the rf-null, the other ion is the only one to possess micromotion sidebands, on which selective quantum control can be performed \cite{leibfried1999individual}. Axial displacement of the ion-crystal is realized by applying a differential voltage on the two endcaps.
In the limit of small amplitudes, the Rabi frequency of the sideband is $\Omega_{mm}=\Omega_0\eta_{mm}$ where $\Omega_0$ is the carrier Rabi frequency, $\eta_{mm}={\bf k}\cdot {\bf x}$ is the micromotion Lamb-Dicke parameter and ${\bf x}$ is the micromotion amplitude along the laser wavevector ${\bf k}$. 
To maintain coherence throughout the experimental sequence, the trap rf and the signals feeding the acousto-optic modulator (AOM) that drive both the micromotion sideband and the carrier transition, are all phase-locked to the same time base.

\begin{figure}[t!]
\centerline{\includegraphics[width=1\columnwidth]{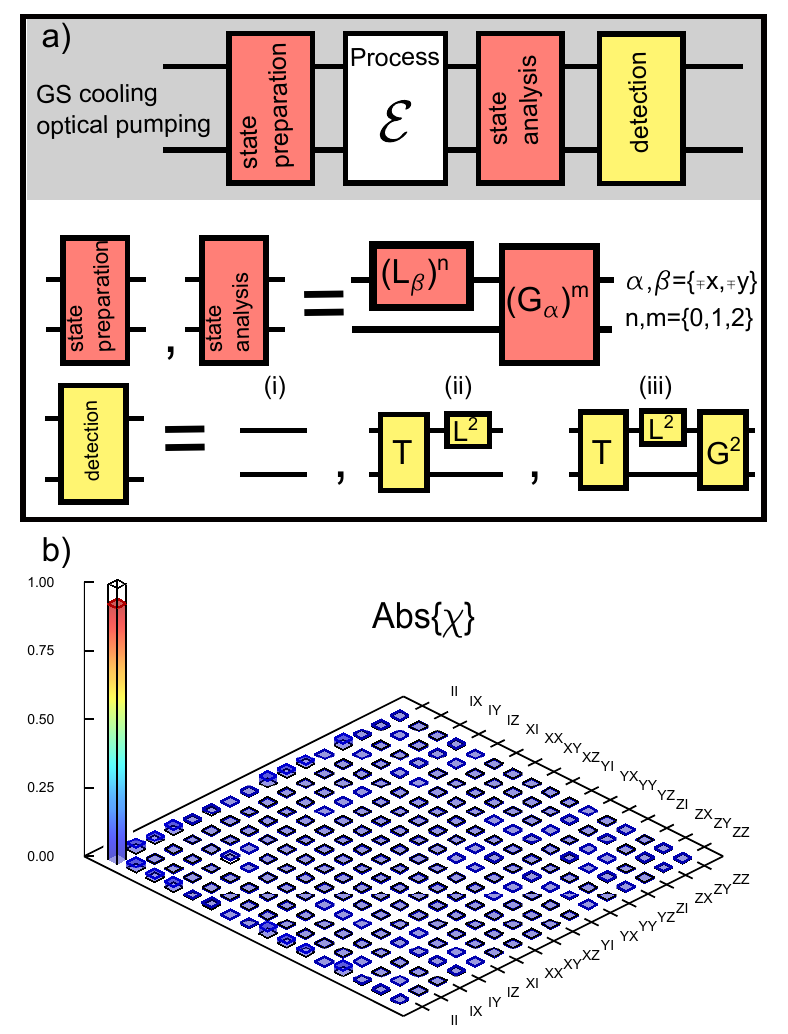}}
\caption{(Color online) Two qubits quantum process tomography. (a) The protocol is composed of products common $\pi/2$ rotations to both qubits ($G$'s) and products of local $\pi/2$ rotations ($L$'s) acting only on the ion with a micromotion sideband. In cases where we need to measure the state of a particular ion the other ion is set to be bright by a global rf $\pi$-pulse, T, that transfers the $\ket{S}$ state population to the state $\ket{S'}$ followed by a another $\pi$-pulse that maps the population in $\ket{D}$ to $\ket{S}$. (b) Absolute value of the $\chi$-matrix of the Identity process ($\chi$ is almost completely real with a maximum imaginary value of 0.02). The basis of the matrix are $\sigma_i\otimes\sigma_j$ where, for clarity we use the convention $X=\sigma_x,Y=\sigma_y,Z=\sigma_z$. The dominant contribution to the process matrix (0.94) is indeed the identity.  
} \label{chiIdentity}
\end{figure}

Our protocol for implementing two qubit QPT is illustrated in Fig.\ref{chiIdentity}a.
In order to measure in all the necessary bases, it is enough to possess single-qubit rotation capability on one qubit only. We look for an operation $R_{ij}$ such that $R_{ij}^\dag(\sigma_i\otimes\sigma_j)R_{ij}=\sigma_z\otimes\sigma_z$. Indeed, $R_{ij}$ can be decomposed in the form: $R_{ij}=G_{\alpha(i,j)}.L_{\beta(i,j)}$, where $G$ ($L$) is a global (local) rotation around a direction lying in the $(x,y)$-plane of the optical qubit. More precisely, these operators can be written as $G_{\alpha}=\exp(i\frac{\pi}{2}(\sigma_{\alpha}^{(1)}+\sigma_{\alpha}^{(2)}))$, a global $\pi/2$ rotation around the $\alpha$-axis, and $L_{\beta}=\exp(i\frac{\pi}{2}\sigma_{\beta}^{(2)})$ a local $\pi/2$ rotation around the $\beta$-axis of only one ion (where $\alpha,\beta=\pm x,\pm y$). For example $R_{xy}=G_{-y}.L_{x}$, $R_{zx}=L_{-y}$, $R_{xz}=G_{-y}.L_y$, and so on. Similarly, the state preparation of all product states mentioned above can be realized using the same set of operations after initializing the ions to $\ket{z}\otimes\ket{z}$ by optical pumping. Lastly, the value of $\braket{\sigma_z\otimes\sigma_z}= P_0+P_2-P_1$ is extracted from fluorescence histograms. In addition, some of the necessary measurements for QPT are of the form $I\otimes\sigma_j$ or $\sigma_j\otimes I$, and thus require the measurement of the state of each ion separately. To perform these measurements we utilize the auxiliary $\ket{S'}$ level to which we transfer one of the ions into a definitely bright state ($\ket{S},\ket{S'}$). This is accomplished by first transferring the $\ket{S}$ state population into $\ket{S'}$ in both ions with an rf $\pi$-pulse. Then another $\pi$-pulse on the micromotion sideband transfers the $\ket{D}$ state population of that ion into $\ket{S}$. The state of the ion at the null is then determined by $P_2$ ($P_0=0$). Similarly, the state of the ion with micromotion is determined by applying an additional global carrier $\pi$-pulse to both ions. \\

Using all the above, we first validate our QPT toolbox by characterizing the identity process, which amounts to concatenating the state preparation and analysis protocols without any intermediate operation. While the quantum state tomography of a two-qubit system requires for each input state 15 independent real-valued parameters (since $\rho$ is hermitian and $\textrm{Tr}[\rho]=1$), a full QPT requires a total $2^{4n}-2^{2n}$ measurements for a system of $n$ qubits. From the measurement of the 16 output density matrices, we reconstruct the $\chi$-matrix \cite{nielsen2010quantum}. Due to noise and systematic errors in the measurements, the algebraically calculated $\chi$-matrix is not physical i.e  does not represent a completely positive map. We obtained a physical $\chi$-matrix by means of maximum likelihood process estimation \cite{Hradil97}.
Fig.\ref{chiIdentity}b. shows the absolute value of the resulting process matrix. Here, the values of the imaginary part are small ($<0.02$). The definition of proper (and simple) distance measures for quantum operations is a subtle problem \cite{Gilchrist05}. For simplicity, we will quantify the proximity between a tomographically reconstructed process $\mathcal{E}_{meas}$ and a target process $\mathcal{E}_0$ by the mean fidelity: $\overline{\mathcal{F}(\mathcal{E}_0[\rho_{in}],\mathcal{E}_{meas}[\rho_{in}])}$, where $\mathcal{F}(\rho,\sigma)=\textrm{Tr}[\sqrt{\sqrt{\rho}\sigma\sqrt{\rho}}]^2$ is the fidelity between the density matrices $\rho$ and $\sigma$, $\rho_{in}=\ket{\Psi_{in}}\bra{\Psi_{in}}$ and the overline indicates average over all possible pure input states $\ket{\Psi_{in}}$. Note that if the target process is unitary, then in the case of a pure input state, the corresponding output state, $\ket{\Psi_{out}}$, is also pure, and therefore the fidelity takes the simpler form $\mathcal{F}(\mathcal{E}_0[\rho_{in}], \mathcal{E}_{meas}[\rho_{in}]) = \braket{\Psi_{out}| \mathcal{E}_{meas}[\rho_{in}]| \Psi_{out}}$. Interestingly, this fidelity (contrary to the trace fidelity \cite{nielsen2010quantum}) is unity if and only if $\mathcal{E}_{meas}=\mathcal{E}_0$ regardless of whether the processes are unitary or not.
For the identity process of Fig.\ref{chiIdentity}, we find a mean fidelity of $0.95(2)$. While a single Rabi flop on the micromotion sideband was performed with a fidelity of $0.99$, slow drifts of stray fields lead to small displacements of the ion from the rf-null. Furthermore small laser detuning errors reduced the single-qubit rotation fidelity.

\begin{figure}[b!]
\centerline{\includegraphics[width=1\columnwidth]{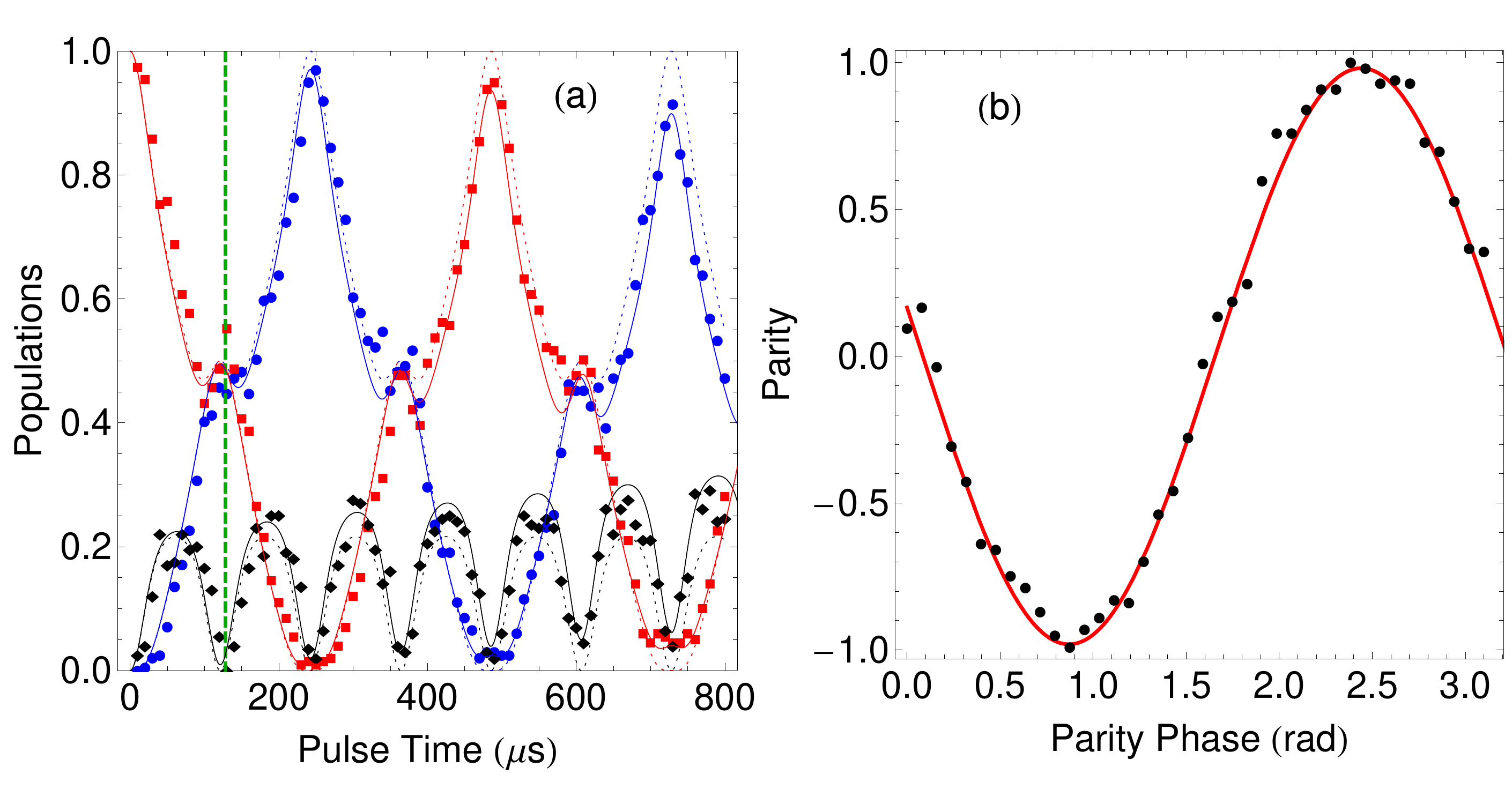}}
\caption{(Color online) A M$\o$lmer-S$\o$rensen entangling gate on a ground-state cooled two-ion chain. (a) Evolution of the populations $P_0$ (blue), $P_1$ (black), and $P_2$ (red) as a function of the interaction duration. The dashed lines are the analytical solutions of the MS model \cite{kirchmair2009deterministic}, the solid lines also take into account depolarization (see text). (b) Parity ($=P_0+P_2-P_1$) oscillation, obtained by scanning the phase of a $\pi/2$-pulse after a gate time of $t_g=130$ $\mu$s. The red solid line is a sine fit to the experimental data. We measure a parity contrast of $0.98(2)$, and $P_1(t_g)=0.01(1)$.
} \label{GatePlots}
\end{figure}
\begin{figure}[t!]
\center
\includegraphics[width=\columnwidth]{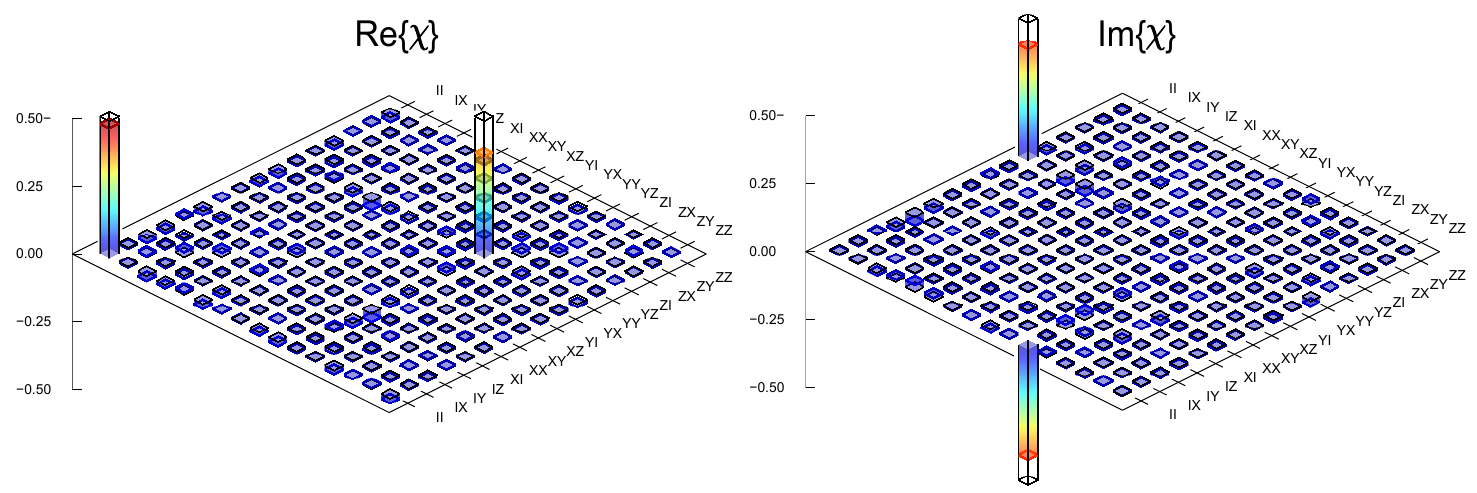}
\includegraphics[width=1\columnwidth]{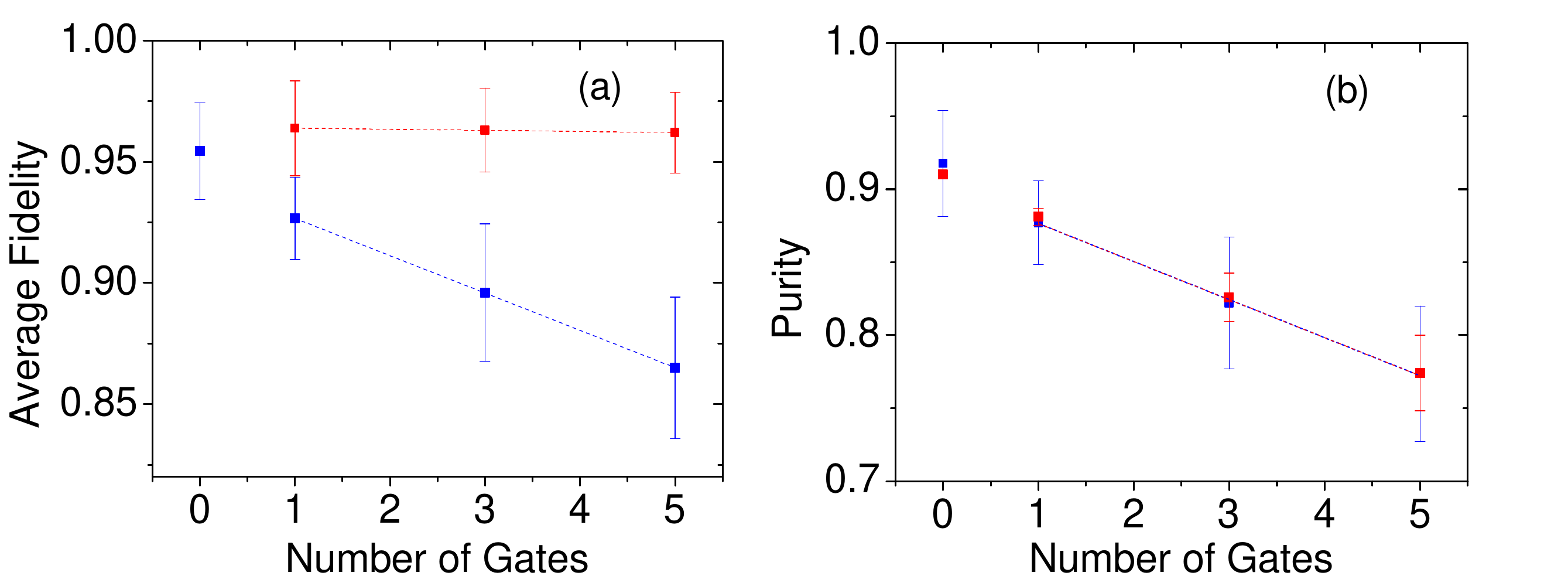}
\caption{(Color online) Process tomography of a M$\o$lmer-S$\o$rensen interaction. Upper panel: Real and imaginary parts of the reconstructed process matrix Re[$\chi$] and Im[$\chi$] of 5 consecutive gates. 
Lower Panel: (a) Average fidelity. In blue points, the values extracted from the experimentally reconstructed $\chi$ matrices with respect to the target gate operation $\mathcal{E}_0$. In red points, the mean values are calculated with respect to output density matrices for which a depolarizing operation has been applied. Dashed lines are linear fits.  (b) Average state purity. In blue points calculated from the reconstructed process $\chi$ matrices and in red from a depolarization channel model with a depolarization rate of $1.8\times 10^{-2}$ per gate time.  
}
\label{AllgatePT}
\end{figure}

Next we apply our tomography protocol to analyze a M$\o$lmer-S$\o$rensen interaction. The gate is performed on the $\ket{S}\rightarrow\ket{D}$ transition via two sidebands, which are generated by applying two rf signals into an acousto-optic modulator (AOM) switch \cite{roos2008ion}, with frequencies of $\omega_c\pm(\delta+\epsilon)$, where $\omega_c$ is the carrier transition frequency, $\delta$ is a motional sideband used for the intermediate spin-motion entanglement and $\epsilon$ is the gate detuning. We use the stretch axial mode at a frequency of $\sqrt{3}\omega_z=1.679$ MHz for entanglement as it is less sensitive to heating than the center of mass mode. The gate detuning is optimally set according to the Rabi frequency, $\Omega$, as $\eta\Omega=\epsilon/4$, where $\eta$ is the motional Lamb-Dicke parameter, and the gate time is $t_g=2\pi/\epsilon$. After ground-state cooling of the stretch mode, the two ions are initialized by optical pumping to $\ket{SS}$. The gate generates the maximally entangled state $\ket{\Phi}=(\ket{SS}+i\ket{DD})/\sqrt{2}$. In Fig.\ref{GatePlots}a, we display the evolution of $P_0$ (blue points), $P_1$(black points), and $P_2$ (red points), obtained for a gate detuning of $\epsilon=2\pi\times7.7$ kHz. At a pulse time of 130 $\mu$s (shown by a vertical dashed green line), the two ions are maximally entangled. Together with parity analysis shown in Fig.\ref{GatePlots}b, we measure a fidelity of the Bell-state production of $\mathcal{F}(\ket{\Phi}\bra{\Phi},\rho_{exp})=0.985(10)$.

After calibrating the gate time, we experimentally reconstructed the $\chi$-process matrices of a single, three and five consecutive MS gates (the latter is shown in the upper panel of Fig.3). Each experiment is repeated 400 times, totaling $240\times400=96000$ measurements for a full process tomography. The target process matrices can be readily deduced from the evolution operator \cite{roos2008ion}. Starting from the hamiltonian describing the ions interacting with the bichromatic field, one can show that at the gate time, the evolution operator reduces to $U_{\textrm{MS}}(t=t_g)=\exp(-i(\pi/8)S_y^2)$. The motional part of the evolution operator is the identity, and only at the gate time (and multiples of it) the internal and motional parts factorize, leading to no loss of coherence for the internal-part due to the tracing of the motional degrees of freedom. The $\chi$-matrix is calculated by expanding the exponent of $U_{\textrm{MS}}$: $U_{\textrm{MS}}=1/\sqrt{2}(I\otimes I+i Y\otimes Y)$. Plugging this expression in Eq.(\ref{chiMatrixDef}), we readily find,
\begin{equation}\label{chiEqMS}
\left\{ \begin{array}{lll}
         (\chi^{MS})_{II,II} &=& (\chi^{MS})_{YY,YY}=1/2\\
         (\chi^{MS})_{II,YY} &=& -(\chi^{MS})_{YY,II}=i/2
\end{array} \right.
\end{equation}
Interestingly, this operation matrix, with four non-zero elements, is considerably simpler in the $\sigma$-operators basis than the previously process-analyzed CNOT gates or iSWAP gates (with sixteen non-null elements each \cite{childs2001realization,bialczak2010quantum}). The dominance of the process matrix elements in Eq.(\ref{chiEqMS}) is in good agreement with our results, shown in Fig.\ref{AllgatePT}.\\

The mean fidelity of $\mathcal{E}_{meas}$ with respect to $\mathcal{E}_0(n)[\rho_{in}] = U^n_\textrm{MS}\rho_{in} U^{\dag n}_\textrm{MS}$ are shown by blue points in Fig.3a. As seen, due to gate imperfections the fidelity decreases with growing number of applied gates at a rate of $1.5\times10^{-2}$ per gate. The direct interpretation of imperfections from the process matrix is notoriously difficult because in the $\sigma$ operators basis each noise process involves multiple elements with various weights. Instead, it is common to compare the measured process to different noise models \cite{childs2001realization,bialczak2010quantum,kofman2009two}. 
We found that the dominant error of our gate is consistent with a quantum depolarization channel for the two ions \cite{nielsen2010quantum}, whose map is $\mathcal{E}_{DC}(t)[\rho]=(1-p(t))\rho+\frac{p(t)}{4} I\otimes I$, where $p(t)=\alpha_{DC}t$, and $\alpha_{DC}$, is the depolarization rate.
The single free parameter of the model $\alpha_{DC}$ can be first determined from the purity of our measured processes, which is affected only by non-unitary operations. We recall that the purity of state $\sigma$ is $\mathcal{P}(\sigma)=\textrm{Tr}[\sigma^2]$. The rate  of depolarization is determined by matching the slope (with respect to the number of gates) of $\overline{\mathcal{P}(\mathcal{E}_{DC}(t)[\rho_\textrm{in}])}$ (red points in Fig.3b) with the experimental purities $\overline{\mathcal{P}(\mathcal{E}_{meas}[\rho_\textrm{in}])}$ (blue points in Fig.3b). The slope matches our data for a rate of $\alpha_{DC}=1.8\times10^{-2}$ per gate time. The identity process is excluded from these fits, since errors from tomography and the gates have a different origin, they are largely independent. 

We can verify the appropriateness of this description by calculating the fidelity of $\mathcal{E}_{meas}$ with respect to an MS interaction that has suffered partial depolarization $\overline{\mathcal{F}(\mathcal{E}_{DC}(nt_g)[\rho_n],\mathcal{E}_{meas}[\rho_\textrm{in}])}$ (red points of Fig.3b), where $\rho_n=U^n_{\textrm{MS}}\rho_\textrm{in}U^{\dag n}_{\textrm{MS}}$ ($n=0,1,3,5$). Remarkably, we find that for the rate previously determined from the averaged purity, the mean fidelity to the partially depolarized states is almost constant as a function of the number of gates applied. This shows that the depolarization channel accounts well for the imperfections introduced by successive applications of gates, and we conclude that the remaining error is due to the tomography itself. 
Morever, we can use the depolarization channel model to predict the expected imperfections on the population time dynamics previously measured in Fig.\ref{GatePlots}. On one hand, these contain much more limited information than the full process matrices, namely only the diagonal elements of the output state starting from $\ket{SS}$, but on the other hand, they are free from tomographic errors. While the solution of the perfect MS propagator (dashed lines in Fig.\ref{GatePlots}) does not describe our data well for the longest times, the agreement is excellent when the depolarization is taken into account (solid lines), especially since there is no adjustable parameter. In particular, at $t=t_g$ we expect a Bell state production fidelity of $98.2$ $\%$, in very good agreement with the experimental determination. While the physical origin of the depolarization is unknown to us, the measured depolarization rate is in rough agreement with off-resonance $S\leftrightarrow D$ incoherent transfer rate we observe on a single trapped ion and that is generated by fast (1 MHz) phase noise of our laser. A more thorough study of the cause for depolarization is under way. 

In conclusion we implemented a simple method for QPT of two-qubit processes based on a single discriminatory transition and with no direct spatially-selective imaging. The protocol was used to tomographically reconstruct a high fidelity M$\o$lmer-S$\o$rensen interaction. The M$\o$lmer-S$\o$rensen interaction is currently the main method for generating entangling gates with trapped ion qubits and for synthesizing coupling between trapped ion spins for quantum simulation and this work provides the first full characterization of its process matrix.

This research was supported by the Israeli Science Foundation, the Minerva Foundation, the German-Israeli Foundation for Scientific Research, the Crown Photonics Center, the Wolfson Family Charitable Trust, Yeda-Sela Center for Basic Research, David Dickstein of France and M. Kushner Schnur, Mexico.


\begin{thebibliography}{24}
\expandafter\ifx\csname natexlab\endcsname\relax\def\natexlab#1{#1}\fi
\expandafter\ifx\csname bibnamefont\endcsname\relax
  \def\bibnamefont#1{#1}\fi
\expandafter\ifx\csname bibfnamefont\endcsname\relax
  \def\bibfnamefont#1{#1}\fi
\expandafter\ifx\csname citenamefont\endcsname\relax
  \def\citenamefont#1{#1}\fi
\expandafter\ifx\csname url\endcsname\relax
  \def\url#1{\texttt{#1}}\fi
\expandafter\ifx\csname urlprefix\endcsname\relax\def\urlprefix{URL }\fi
\providecommand{\bibinfo}[2]{#2}
\providecommand{\eprint}[2][]{\url{#2}}

\bibitem[{\citenamefont{Barenco et~al.}(1995)\citenamefont{Barenco, Bennett,
  Cleve, DiVincenzo, Margolus, Shor, Sleator, Smolin, and
  Weinfurter}}]{barenco1995elementary}
\bibinfo{author}{\bibfnamefont{A.}~\bibnamefont{Barenco}},
  \bibinfo{author}{\bibfnamefont{C.}~\bibnamefont{Bennett}},
  \bibinfo{author}{\bibfnamefont{R.}~\bibnamefont{Cleve}},
  \bibinfo{author}{\bibfnamefont{D.}~\bibnamefont{DiVincenzo}},
  \bibinfo{author}{\bibfnamefont{N.}~\bibnamefont{Margolus}},
  \bibinfo{author}{\bibfnamefont{P.}~\bibnamefont{Shor}},
  \bibinfo{author}{\bibfnamefont{T.}~\bibnamefont{Sleator}},
  \bibinfo{author}{\bibfnamefont{J.}~\bibnamefont{Smolin}}, \bibnamefont{and}
  \bibinfo{author}{\bibfnamefont{H.}~\bibnamefont{Weinfurter}},
  \bibinfo{journal}{Phys. Rev. A} \textbf{\bibinfo{volume}{52}},
  \bibinfo{pages}{3457} (\bibinfo{year}{1995}).

\bibitem[{\citenamefont{Kiesel et~al.}(2005)\citenamefont{Kiesel, Schmid,
  Weber, Ursin, and Weinfurter}}]{kiesel2005linear}
\bibinfo{author}{\bibfnamefont{N.}~\bibnamefont{Kiesel}},
  \bibinfo{author}{\bibfnamefont{C.}~\bibnamefont{Schmid}},
  \bibinfo{author}{\bibfnamefont{U.}~\bibnamefont{Weber}},
  \bibinfo{author}{\bibfnamefont{R.}~\bibnamefont{Ursin}}, \bibnamefont{and}
  \bibinfo{author}{\bibfnamefont{H.}~\bibnamefont{Weinfurter}},
  \bibinfo{journal}{Phys. Rev. Lett.} \textbf{\bibinfo{volume}{95}},
  \bibinfo{pages}{210505} (\bibinfo{year}{2005}).

\bibitem[{\citenamefont{Childs et~al.}(2001)\citenamefont{Childs, Chuang, and
  Leung}}]{childs2001realization}
\bibinfo{author}{\bibfnamefont{A.}~\bibnamefont{Childs}},
  \bibinfo{author}{\bibfnamefont{I.}~\bibnamefont{Chuang}}, \bibnamefont{and}
  \bibinfo{author}{\bibfnamefont{D.}~\bibnamefont{Leung}},
  \bibinfo{journal}{Phys. Rev. A} \textbf{\bibinfo{volume}{64}},
  \bibinfo{pages}{012314} (\bibinfo{year}{2001}).

\bibitem[{\citenamefont{Riebe et~al.}(2006)\citenamefont{Riebe, Kim, Schindler,
  Monz, Schmidt, Korber, Hansel, Haffner, Roos, and Blatt}}]{riebe2006process}
\bibinfo{author}{\bibfnamefont{M.}~\bibnamefont{Riebe}},
  \bibinfo{author}{\bibfnamefont{K.}~\bibnamefont{Kim}},
  \bibinfo{author}{\bibfnamefont{P.}~\bibnamefont{Schindler}},
  \bibinfo{author}{\bibfnamefont{T.}~\bibnamefont{Monz}},
  \bibinfo{author}{\bibfnamefont{P.}~\bibnamefont{Schmidt}},
  \bibinfo{author}{\bibfnamefont{T.}~\bibnamefont{Korber}},
  \bibinfo{author}{\bibfnamefont{W.}~\bibnamefont{Hansel}},
  \bibinfo{author}{\bibfnamefont{H.}~\bibnamefont{Haffner}},
  \bibinfo{author}{\bibfnamefont{C.}~\bibnamefont{Roos}}, \bibnamefont{and}
  \bibinfo{author}{\bibfnamefont{R.}~\bibnamefont{Blatt}},
  \bibinfo{journal}{Phys. Rev. Lett.} \textbf{\bibinfo{volume}{97}},
  \bibinfo{pages}{220407} (\bibinfo{year}{2006}).

\bibitem[{\citenamefont{Home et~al.}(2009)\citenamefont{Home, Hanneke, Jost,
  Amini, Leibfried, and Wineland}}]{home2009complete}
\bibinfo{author}{\bibfnamefont{J.}~\bibnamefont{Home}},
  \bibinfo{author}{\bibfnamefont{D.}~\bibnamefont{Hanneke}},
  \bibinfo{author}{\bibfnamefont{J.}~\bibnamefont{Jost}},
  \bibinfo{author}{\bibfnamefont{J.}~\bibnamefont{Amini}},
  \bibinfo{author}{\bibfnamefont{D.}~\bibnamefont{Leibfried}},
  \bibnamefont{and} \bibinfo{author}{\bibfnamefont{D.}~\bibnamefont{Wineland}},
  \bibinfo{journal}{Science} \textbf{\bibinfo{volume}{325}},
  \bibinfo{pages}{1227} (\bibinfo{year}{2009}).

\bibitem[{\citenamefont{Bialczak et~al.}(2010)\citenamefont{Bialczak, Ansmann,
  Hofheinz, Lucero, Neeley, O'Connell, Sank, Wang, Wenner, Steffen
  et~al.}}]{bialczak2010quantum}
\bibinfo{author}{\bibfnamefont{R.}~\bibnamefont{Bialczak}},
  \bibinfo{author}{\bibfnamefont{M.}~\bibnamefont{Ansmann}},
  \bibinfo{author}{\bibfnamefont{M.}~\bibnamefont{Hofheinz}},
  \bibinfo{author}{\bibfnamefont{E.}~\bibnamefont{Lucero}},
  \bibinfo{author}{\bibfnamefont{M.}~\bibnamefont{Neeley}},
  \bibinfo{author}{\bibfnamefont{A.}~\bibnamefont{O'Connell}},
  \bibinfo{author}{\bibfnamefont{D.}~\bibnamefont{Sank}},
  \bibinfo{author}{\bibfnamefont{H.}~\bibnamefont{Wang}},
  \bibinfo{author}{\bibfnamefont{J.}~\bibnamefont{Wenner}}, \bibnamefont{and}
  \bibinfo{author}{\bibfnamefont{M.}~\bibnamefont{Steffen}},
  \bibinfo{journal}{Nat. Phys.} \textbf{\bibinfo{volume}{6}}, \bibinfo{pages}{409} (\bibinfo{year}{2010}).

\bibitem[{\citenamefont{S{\o}rensen and M{\o}lmer}(1999)}]{sorensen1999quantum}
\bibinfo{author}{\bibfnamefont{A.}~\bibnamefont{S{\o}rensen}} \bibnamefont{and}
  \bibinfo{author}{\bibfnamefont{K.}~\bibnamefont{M{\o}lmer}},
  \bibinfo{journal}{Phys. Rev. Lett.} \textbf{\bibinfo{volume}{82}},
  \bibinfo{pages}{1971} (\bibinfo{year}{1999}).

\bibitem[{\citenamefont{Sackett et~al.}(2000)\citenamefont{Sackett, Kielpinski,
  King, Langer, Meyer, Myatt, Rowe, Turchette, Itano, Wineland
  et~al.}}]{sackett2000experimental}
\bibinfo{author}{\bibfnamefont{C.}~\bibnamefont{Sackett}},
  \bibinfo{author}{\bibfnamefont{D.}~\bibnamefont{Kielpinski}},
  \bibinfo{author}{\bibfnamefont{B.}~\bibnamefont{King}},
  \bibinfo{author}{\bibfnamefont{C.}~\bibnamefont{Langer}},
  \bibinfo{author}{\bibfnamefont{V.}~\bibnamefont{Meyer}},
  \bibinfo{author}{\bibfnamefont{C.}~\bibnamefont{Myatt}},
  \bibinfo{author}{\bibfnamefont{M.}~\bibnamefont{Rowe}},
  \bibinfo{author}{\bibfnamefont{Q.}~\bibnamefont{Turchette}},
  \bibinfo{author}{\bibfnamefont{W.}~\bibnamefont{Itano}}, \bibnamefont{and}
  \bibinfo{author}{\bibfnamefont{D.}~\bibnamefont{Wineland}}, 
  \bibinfo{journal}{Nature} \textbf{\bibinfo{volume}{404}}, \bibinfo{pages}{256} (\bibinfo{year}{2000}).

\bibitem[{\citenamefont{Leibfried et~al.}(2003)\citenamefont{Leibfried,
  DeMarco, Meyer, Lucas, Barrett, Britton, Itano, Jelenkovic, Langer, Rosenband
  et~al.}}]{leibfried2003experimental}
\bibinfo{author}{\bibfnamefont{D.}~\bibnamefont{Leibfried}},
  \bibinfo{author}{\bibfnamefont{B.}~\bibnamefont{DeMarco}},
  \bibinfo{author}{\bibfnamefont{V.}~\bibnamefont{Meyer}},
  \bibinfo{author}{\bibfnamefont{D.}~\bibnamefont{Lucas}},
  \bibinfo{author}{\bibfnamefont{M.}~\bibnamefont{Barrett}},
  \bibinfo{author}{\bibfnamefont{J.}~\bibnamefont{Britton}},
  \bibinfo{author}{\bibfnamefont{W.}~\bibnamefont{Itano}},
  \bibinfo{author}{\bibfnamefont{B.}~\bibnamefont{Jelenkovic}},
  \bibinfo{author}{\bibfnamefont{C.}~\bibnamefont{Langer}}, \bibnamefont{and}
  \bibinfo{author}{\bibfnamefont{T.}~\bibnamefont{Rosenband}},
  \bibinfo{journal}{Nature} \textbf{\bibinfo{volume}{422}}, \bibinfo{pages}{412} (\bibinfo{year}{2003}).

\bibitem[{\citenamefont{Kim et~al.}(2009)\citenamefont{Kim, Chang, Islam,
  Korenblit, Duan, and Monroe}}]{kim2009entanglement}
\bibinfo{author}{\bibfnamefont{K.}~\bibnamefont{Kim}},
  \bibinfo{author}{\bibfnamefont{M.}~\bibnamefont{Chang}},
  \bibinfo{author}{\bibfnamefont{R.}~\bibnamefont{Islam}},
  \bibinfo{author}{\bibfnamefont{S.}~\bibnamefont{Korenblit}},
  \bibinfo{author}{\bibfnamefont{L.}~\bibnamefont{Duan}}, \bibnamefont{and}
  \bibinfo{author}{\bibfnamefont{C.}~\bibnamefont{Monroe}},
  \bibinfo{journal}{Phys. Rev. Lett.} \textbf{\bibinfo{volume}{103}},
  \bibinfo{pages}{120502} (\bibinfo{year}{2009}).

\bibitem[{\citenamefont{Benhelm et~al.}(2008)\citenamefont{Benhelm, Kirchmair,
  Roos, and Blatt}}]{benhelm2008towards}
\bibinfo{author}{\bibfnamefont{J.}~\bibnamefont{Benhelm}},
  \bibinfo{author}{\bibfnamefont{G.}~\bibnamefont{Kirchmair}},
  \bibinfo{author}{\bibfnamefont{C.}~\bibnamefont{Roos}}, \bibnamefont{and}
  \bibinfo{author}{\bibfnamefont{R.}~\bibnamefont{Blatt}},
  \bibinfo{journal}{Nat. Phys.} \textbf{\bibinfo{volume}{4}},
  \bibinfo{pages}{463} (\bibinfo{year}{2008}).

\bibitem[{\citenamefont{Kirchmair et~al.}(2009)\citenamefont{Kirchmair,
  Benhelm, Z{\"a}hringer, Gerritsma, Roos, and
  Blatt}}]{kirchmair2009deterministic}
\bibinfo{author}{\bibfnamefont{G.}~\bibnamefont{Kirchmair}},
  \bibinfo{author}{\bibfnamefont{J.}~\bibnamefont{Benhelm}},
  \bibinfo{author}{\bibfnamefont{F.}~\bibnamefont{Z{\"a}hringer}},
  \bibinfo{author}{\bibfnamefont{R.}~\bibnamefont{Gerritsma}},
  \bibinfo{author}{\bibfnamefont{C.}~\bibnamefont{Roos}}, \bibnamefont{and}
  \bibinfo{author}{\bibfnamefont{R.}~\bibnamefont{Blatt}},
  \bibinfo{journal}{New J. Phys.} \textbf{\bibinfo{volume}{11}},
  \bibinfo{pages}{023002} (\bibinfo{year}{2009}).

\bibitem[{\citenamefont{Monz et~al.}(2011)\citenamefont{Monz, Schindler,
  Barreiro, Chwalla, Nigg, Coish, Harlander, H{\"a}nsel, Hennrich, and
  Blatt}}]{monz201114}
\bibinfo{author}{\bibfnamefont{T.}~\bibnamefont{Monz}},
  \bibinfo{author}{\bibfnamefont{P.}~\bibnamefont{Schindler}},
  \bibinfo{author}{\bibfnamefont{J.}~\bibnamefont{Barreiro}},
  \bibinfo{author}{\bibfnamefont{M.}~\bibnamefont{Chwalla}},
  \bibinfo{author}{\bibfnamefont{D.}~\bibnamefont{Nigg}},
  \bibinfo{author}{\bibfnamefont{W.}~\bibnamefont{Coish}},
  \bibinfo{author}{\bibfnamefont{M.}~\bibnamefont{Harlander}},
  \bibinfo{author}{\bibfnamefont{W.}~\bibnamefont{H{\"a}nsel}},
  \bibinfo{author}{\bibfnamefont{M.}~\bibnamefont{Hennrich}}, \bibnamefont{and}
  \bibinfo{author}{\bibfnamefont{R.}~\bibnamefont{Blatt}},
  \bibinfo{journal}{Phys. Rev. Lett.} \textbf{\bibinfo{volume}{106}},
  \bibinfo{pages}{130506} (\bibinfo{year}{2011}).

\bibitem[{\citenamefont{Turchette et~al.}(1998)\citenamefont{Turchette, Wood,
  King, Myatt, Leibfried, Itano, Monroe, and
  Wineland}}]{turchette1998deterministic}
\bibinfo{author}{\bibfnamefont{Q.}~\bibnamefont{Turchette}},
  \bibinfo{author}{\bibfnamefont{C.}~\bibnamefont{Wood}},
  \bibinfo{author}{\bibfnamefont{B.}~\bibnamefont{King}},
  \bibinfo{author}{\bibfnamefont{C.}~\bibnamefont{Myatt}},
  \bibinfo{author}{\bibfnamefont{D.}~\bibnamefont{Leibfried}},
  \bibinfo{author}{\bibfnamefont{W.}~\bibnamefont{Itano}},
  \bibinfo{author}{\bibfnamefont{C.}~\bibnamefont{Monroe}}, \bibnamefont{and}
  \bibinfo{author}{\bibfnamefont{D.}~\bibnamefont{Wineland}},
  \bibinfo{journal}{Phys. Rev. Lett.} \textbf{\bibinfo{volume}{81}},
  \bibinfo{pages}{3631} (\bibinfo{year}{1998}).

\bibitem[{\citenamefont{Warring et~al.}(2012)\citenamefont{Warring, Ospelkaus,
  Colombe, Jordens, Leibfried, and Wineland}}]{warring2012individual}
\bibinfo{author}{\bibfnamefont{U.}~\bibnamefont{Warring}},
  \bibinfo{author}{\bibfnamefont{C.}~\bibnamefont{Ospelkaus}},
  \bibinfo{author}{\bibfnamefont{Y.}~\bibnamefont{Colombe}},
  \bibinfo{author}{\bibfnamefont{R.}~\bibnamefont{Jordens}},
  \bibinfo{author}{\bibfnamefont{D.}~\bibnamefont{Leibfried}},
  \bibnamefont{and} \bibinfo{author}{\bibfnamefont{D.}~\bibnamefont{Wineland}},
  \bibinfo{journal}{Phys. Rev. Lett.} \textbf{\bibinfo{volume}{110}},
  \bibinfo{pages}{173002} (\bibinfo{year}{2013}).

\bibitem[{\citenamefont{Navon et~al.}(2012)\citenamefont{Navon, Kotler,
  Akerman, Glickman, Almog, and Ozeri}}]{navon2012single}
\bibinfo{author}{\bibfnamefont{N.}~\bibnamefont{Navon}},
  \bibinfo{author}{\bibfnamefont{S.}~\bibnamefont{Kotler}},
  \bibinfo{author}{\bibfnamefont{N.}~\bibnamefont{Akerman}},
  \bibinfo{author}{\bibfnamefont{Y.}~\bibnamefont{Glickman}},
  \bibinfo{author}{\bibfnamefont{I.}~\bibnamefont{Almog}}, \bibnamefont{and}
  \bibinfo{author}{\bibfnamefont{R.}~\bibnamefont{Ozeri}},
  \bibinfo{journal}{Phys. Rev. Lett.} \textbf{\bibinfo{volume}{111}},
  \bibinfo{pages}{073001} (\bibinfo{year}{2013}).

\bibitem[{\citenamefont{Poyatos et~al.}(1997)\citenamefont{Poyatos, Cirac, and
  Zoller}}]{poyatos1997complete}
\bibinfo{author}{\bibfnamefont{J.}~\bibnamefont{Poyatos}},
  \bibinfo{author}{\bibfnamefont{J.}~\bibnamefont{Cirac}}, \bibnamefont{and}
  \bibinfo{author}{\bibfnamefont{P.}~\bibnamefont{Zoller}},
  \bibinfo{journal}{Phys. Rev. Lett.} \textbf{\bibinfo{volume}{78}},
  \bibinfo{pages}{390} (\bibinfo{year}{1997}).

\bibitem[{\citenamefont{Nielsen and Chuang}(2010)}]{nielsen2010quantum}
\bibinfo{author}{\bibfnamefont{M.}~\bibnamefont{Nielsen}} \bibnamefont{and}
  \bibinfo{author}{\bibfnamefont{I.}~\bibnamefont{Chuang}},
  \emph{\bibinfo{title}{Quantum computation and quantum information}}
  (\bibinfo{publisher}{Cambridge University Press}, \bibinfo{year}{2010}).

\bibitem[{\citenamefont{Akerman et~al.}(2011)\citenamefont{Akerman, Glickman,
  Kotler, Keselman, and Ozeri}}]{akerman2011quantum}
\bibinfo{author}{\bibfnamefont{N.}~\bibnamefont{Akerman}},
  \bibinfo{author}{\bibfnamefont{Y.}~\bibnamefont{Glickman}},
  \bibinfo{author}{\bibfnamefont{S.}~\bibnamefont{Kotler}},
  \bibinfo{author}{\bibfnamefont{A.}~\bibnamefont{Keselman}}, \bibnamefont{and}
  \bibinfo{author}{\bibfnamefont{R.}~\bibnamefont{Ozeri}},
  \bibinfo{journal}{Appl. Phys. B} \textbf{\bibinfo{volume}{107}},
  \bibinfo{pages}{1167} (\bibinfo{year}{2012}).

\bibitem[{\citenamefont{Letchumanan et~al.}(2005)\citenamefont{Letchumanan, Wilson,
  Gill, Sinclair}}]{letchumanan2005lifetime}
\bibinfo{author}{\bibfnamefont{V.}~\bibnamefont{Letchumanan}},
  \bibinfo{author}{\bibfnamefont{Y.}~\bibnamefont{Wilson}},
  \bibinfo{author}{\bibfnamefont{M.A.}~\bibnamefont{Gill}},\bibnamefont{and}
  \bibinfo{author}{\bibfnamefont{A.G.}~\bibnamefont{Sinclair}},
  \bibinfo{journal}{Phys. Rev. A} \textbf{\bibinfo{volume}{72}},
  \bibinfo{pages}{012509} (\bibinfo{year}{2005}).

\bibitem[{Dio()}]{DiodePaperInPrep}
\bibinfo{note}{N. Akerman {\it et. al.}, in preparation.}

\bibitem[{\citenamefont{Leibfried}(1999)}]{leibfried1999individual}
\bibinfo{author}{\bibfnamefont{D.}~\bibnamefont{Leibfried}},
  \bibinfo{journal}{Phys. Rev. A} \textbf{\bibinfo{volume}{60}},
  \bibinfo{pages}{3335} (\bibinfo{year}{1999}).

\bibitem[{\citenamefont{Hradil}(1997)\citenamefont{Hradil}}]{Hradil97}
\bibinfo{author}{\bibfnamefont{Z.}~\bibnamefont{Hradil}},
  \bibinfo{journal}{Phys. Rev. A} \textbf{\bibinfo{volume}{55}},
  \bibinfo{pages}{R1561} (\bibinfo{year}{1997}).
	
	\bibitem[{\citenamefont{Gilchrist}(2005)\citenamefont{Gilchrist}}]{Gilchrist05}
\bibinfo{author}{\bibfnamefont{A.}~\bibnamefont{Gilchrist}},
\bibinfo{author}{\bibfnamefont{N.}~\bibnamefont{Langford}},
\bibinfo{author}{\bibfnamefont{M.}~\bibnamefont{Nielsen}},
  \bibinfo{journal}{Phys. Rev. A} \textbf{\bibinfo{volume}{71}},
  \bibinfo{pages}{062310} (\bibinfo{year}{2005}).

\bibitem[{\citenamefont{Roos}(2008)}]{roos2008ion}
\bibinfo{author}{\bibfnamefont{C.}~\bibnamefont{Roos}}, \bibinfo{journal}{New
  J. Phys.} \textbf{\bibinfo{volume}{10}}, \bibinfo{pages}{013002}
  (\bibinfo{year}{2008}).	
	
\bibitem[{\citenamefont{Kofman}(2009)}]{kofman2009two}
\bibinfo{author}{\bibfnamefont{A.G.}~\bibnamefont{Kofman}} \bibnamefont{and}
  \bibinfo{author}{\bibfnamefont{A.N.}~\bibnamefont{Korotkov}},
  \bibinfo{journal}{Phys. Rev. A} \textbf{\bibinfo{volume}{80}},
  \bibinfo{pages}{042103} (\bibinfo{year}{2009}).

\end{thebibliography}
\end{document}